\documentclass[12pt,preprint]{aastex}
\def\wise{{\it WISE }}

\begin{document}

\shortauthors{Luhman et al.}
\shorttitle{New MLT Companions}

\title{New M, L, and T Dwarf Companions to Nearby Stars from the Wide-field
Infrared Survey Explorer\altaffilmark{1}}

\author{
Kevin L. Luhman\altaffilmark{2,3},
Nicholas P. Loutrel\altaffilmark{2},
Nicholas S. McCurdy\altaffilmark{2},
Gregory N. Mace\altaffilmark{4},
Nicole D. Melso\altaffilmark{2},
Kimberly M. Star\altaffilmark{2},
Michael D. Young\altaffilmark{5},
Ryan C. Terrien\altaffilmark{2},
Ian S. McLean\altaffilmark{4},
J. Davy Kirkpatrick\altaffilmark{6},
Katherine L. Rhode\altaffilmark{5}
}

\altaffiltext{1}
{Based on data from the {\it Wide-field Infrared Survey Explorer}, 2MASS,
the W.M. Keck Observatory, the NASA Infrared Telescope Facility, the
Hobby-Eberly Telescope, the WIYN Observatory at Kitt Peak National Observatory, 
the {\it Spitzer Space Telescope}, the Canada-France-Hawaii Telescope,
and the European Southern Observatory New Technology Telescope.}

\altaffiltext{2}{Department of Astronomy and Astrophysics,
The Pennsylvania State University, University Park, PA 16802, USA;
kluhman@astro.psu.edu}
\altaffiltext{3}{Center for Exoplanets and Habitable Worlds, The 
Pennsylvania State University, University Park, PA 16802, USA}
\altaffiltext{4}{UCLA Division of Astronomy and Astrophysics, Los Angeles,
CA 90095, USA}
\altaffiltext{5}{Department of Astronomy, Indiana University, Swain West 319,
727 East Third Street, Bloomington, IN 47405, USA}
\altaffiltext{6}{Infrared Processing and Analysis Center, MS 100-22, California
Institute of Technology, Pasadena, CA 91125, USA}

\begin{abstract}

We present 11 candidate late-type companions to nearby stars identified with
data from the {\it Wide-field Infrared Survey Explorer (WISE)} and the
Two Micron All-Sky Survey (2MASS).
Eight of the candidates are likely to be companions based on their
common proper motions with the primaries. The remaining three objects
are rejected as companions, one of which is a free-floating T7 dwarf.
Spectral types are available for five of the companions, which consist
of M2V, M8.5V, L5, T8, and T8. Based on their photometry, the unclassified
companions are probably two mid-M dwarfs and one late-M/early-L dwarf.
One of the T8 companions, WISE J142320.84+011638.0, has already been reported
by Pinfield and coworkers. The other T8 companion, ULAS J095047.28+011734.3,
was discovered by Burningham and coworkers through the United Kingdom Infrared
Telescope Infrared Deep Sky Survey, but its companionship has not been
previously recognized in the literature.
The L5 companion, 2MASS J17430860+8526594, is a new member
of a class of L dwarfs that exhibit unusually blue near-IR colors.
Among the possible mechanisms that have been previously proposed for
the peculiar colors of these L dwarfs, low metallicity does not appear
to be a viable explanation for 2MASS J17430860+8526594 since our
spectrum of the primary suggests that its metallicity is not
significantly subsolar.

\end{abstract}


\section{Introduction}

Among the hundreds of known L and T dwarfs in the solar neighborhood, the
few dozen that reside in multiple systems with stars have the most
accurately characterized physical properties because their ages and
metallicities are relatively well-constrained via their primaries.
As a result, late-type companions to nearby stars have been highly
sought after for testing models of the atmospheres and interiors of
low-mass stars and brown dwarfs \citep{bar03,all11,bur11,sau12}.
Candidate companions have been initially
identified through their colors and/or proper motions,
which are measured through observations that target nearby
stars as well as wide-field surveys like
the Two Micron All-Sky Survey \citep[2MASS,][]{skr06},
the Sloan Digital Sky Survey \cite[SDSS,][]{yor00},
the SuperCOSMOS Sky Survey \citep[SSS,][]{ham01},
and the United Kingdom Infrared Telescope Infrared Deep Sky Survey
\citep[UKIDSS,][]{law07}.
More specifically, companion surveys have utilized near-infrared (IR) colors
from targeted imaging \citep{bec88} and 2MASS \citep{giz01},
proper motions from seeing-limited \citep{mug06}
and adaptive optics images \citep{nak95,pot02,met06,tha09},
narrow-band colors obtained with adaptive optics \citep{bil06}, and
mid-IR colors \citep{luh07} and proper motions \citep{luh11} from
the {\it Spitzer Space Telescope} \citep{wer04}.
Additional companions have been found during surveys for
free-floating L and T dwarfs, and later recognized as comoving with nearby
stars \citep{fah10}. Those surveys identified L/T candidates based on 
2MASS colors \citep{bur00,kir01,bur05} and proper motions \citep{wil01},
near-IR colors from UKIDSS \citep{burn09,gol10,day11},
optical colors from SDSS \citep{zha10},
proper motions from UKIDSS and SDSS \citep{sch10},
and proper motions from SSS \citep{sch03}.

The {\it Wide-field Infrared Survey Explorer} \citep[{\it WISE},][]{wri10}
and Pan-STARRS1 \citep[PS1,][]{kai02} are two of the
newest resources for uncovering late-type companions in the solar neighborhood.
Because \wise imaged the entire sky at mid-IR wavelengths, it is the best
available tool for a wide-field census of the coolest nearby brown dwarfs
\citep{kir11,cus11}.
PS1 does not reach the temperature limits of {\it WISE} since
it operates at optical and near-IR wavelengths, but 
its combination of area, depth, spatial resolution, and temporal sampling 
greatly enhances the potential for proper motion surveys \citep{dea11,liu11}.
Late-type companions already have been discovered using 
colors from UKIDSS and \wise \citep{pin12} as well as
proper motions from 2MASS and \wise \citep{lou11,muz12}
and 2MASS and PS1 \citep{dea12a,dea12b}.
In this paper, we present the latest discoveries from our ongoing survey
for late-type companions with 2MASS and {\it WISE}.

\section{Search Methods}
\label{sec:methods}

The primary sources of data for our survey for companions are the
2MASS Point Source Catalog and the {\it WISE} All-Sky Source Catalog.
The images from 2MASS were collected between 1997 and 2001
in broad-band filters centered at 1.25, 1.65, and 2.16~\micron\ \citep[$J$,
$H$, $K_s$,][]{skr06}. The {\it WISE} images were obtained during 2010
through filters at 3.4, 4.6, 12, and 22~\micron\ \citep[W1--W4,][]{wri10}.

Our companion survey consists of two complementary strategies.
In the first one, we are searching for proper motion companions to stars
and brown dwarfs within $\sim30$~pc \citep[e.g.,][]{sub05,lep05a,lep08,fah09}
using astrometry from 2MASS and {\it WISE}. 
We discovered an L dwarf companion with this method shortly after {\it WISE}
data became publicly available \citep{lou11}. Details of the data analysis
are described in that study.

In our second survey strategy, we are searching for {\it WISE} sources that lack
2MASS counterparts and that exhibit {\it WISE} colors and magnitudes that are
consistent with those expected for cool companions to nearby stars
\citep{kir11,kir12}. The photometric selection criteria consist of
the following: $\sigma(W2)<0.1$, $W2-W3<2.5$ if $\sigma(W3)<0.25$, $W1-W2>1$
($\gtrsim$T1), and $2\times W2-W1<DM+11.5$, where DM is the distance modulus
of the primary, which is computed from the parallax when available, and
the photometric distance otherwise. In addition to {\it WISE} sources that have
no 2MASS counterparts within $3\arcsec$, we include {\it WISE} sources that
are detected by 2MASS but that have uncertain photometry ($>$0.1~mag) in all
2MASS bands since they are excluded from the 2MASS/{\it WISE} proper motion
search.  For each candidate identified in this
manner within a projected separation of 10,000~AU from the primary, we verify
that it is unaffected by artifacts or blends in the {\it WISE} images and
is not detected in optical images from the Digitized Sky Survey.
We also attempt to measure the proper motion and further constrain
the colors of each candidate by checking for detections in images from
wide-field surveys (e.g., UKIDSS) and the public data archives of observatories.
If a candidate is sufficiently promising, we pursue followup imaging
or spectroscopy to verify its cool nature and its companionship.
For these followup observations, we give higher priority to candidates
with smaller separations from the primary stars since they are more likely
to be bona fide companions.

\begin{figure}
\epsscale{1}
\plotone{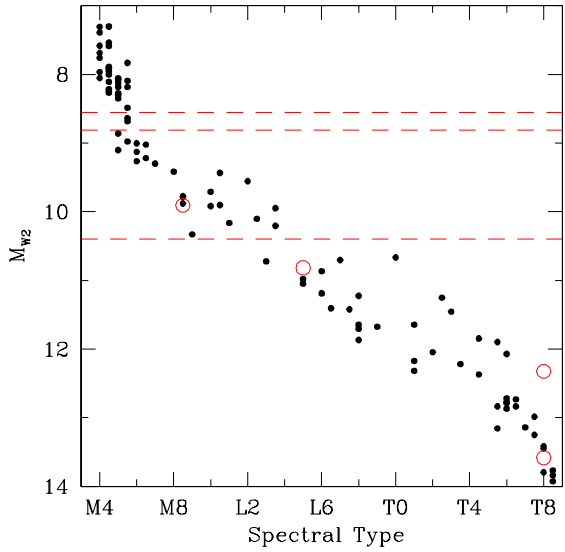}
\caption{
$M_{W2}$ versus spectral type for the candidate companions that exhibit
common proper motions with their primaries (open circles and dashed lines, 
Table~\ref{tab:comp}) and a sample of M, L, and T dwarfs with measured
distances \citep[filled circles,][http://DwarfArchives.org]{leg10a,kir11,kir12}.
}
\label{fig:spw2}
\end{figure}

In the next section, we describe our new sample of 11 candidate companions
to nearby stars. Table~\ref{tab:comp} lists the spectral types, proper motions,
and distances of the primaries and spectral types, proper motions, and
separations of the candidates. 
The 2MASS and {\it WISE} photometric data for the candidates are given in
Table~\ref{tab:phot}.
Although we are primarily interested in finding L and T dwarf companions,
we note the new M dwarf companions that have been uncovered as well.
All candidates except the earliest one are plotted in color-magnitude and
color-color
diagrams in Figure~\ref{fig:cmd}, and those that are confirmed as companions
based on their proper motions (see Table~\ref{tab:comp}) are shown in
a diagram of absolute magnitude versus spectral type in Figure~\ref{fig:spw2}.
For reference, these diagrams also include samples of known M, L, and T dwarfs.
The M dwarfs are from the compilation of stars within 8~pc by \citet{kir12}
and the L and T dwarfs are from the compilations by \citet{leg10a},
\citet{kir11}, and http://DwarfArchives.org.
Because the $J$-band data for our candidate T dwarfs have been measured with
the Mauna Kea Observatories (MKO) system, we adopt $J$-band data from that
system for the known T dwarfs as well. For the known M and L dwarfs, we adopt
$J$, $H$, and $K_s$ from 2MASS.

\section{Candidate Companions}

\subsection{2MASS 03184214+0828002}

Through our 2MASS/{\it WISE} proper motion analysis, we identified 
2MASS~J03184214+0828002 (hereafter 2M~0318+0828) as a possible companion
to the $6\arcsec$ binary system LSPM~J0318+0827S/N \citep{lep05b}.
The brighter component of the binary, LSPM~J0318+0827S, has a spectral type of
M3. LSPM~J0318+0827N is unclassified, but it is probably later than the primary
by $\sim1$~subclass based on the relative fluxes of the pair
($\Delta K_s=1.0$~mag).
We estimated a distance of 23~pc for LSPM~J0318+0827S by combining its
2MASS photometry with the typical absolute magnitudes for an M3 dwarf
\citep{kra07}. At this distance, the $75\arcsec$ angular separation
between 2M~0318+0828 and LSPM~J0318+0827S corresponds to 1700~AU.
The proper motion of 2M~0318+0828 measured from 2MASS and {\it WISE} agrees
with that of LSPM~J0318+0827S/N (Table~\ref{tab:comp}), i.e., their 1~$\sigma$
errors overlap.
The colors and magnitudes of 2M~0318+0828 are consistent with those
expected for a late-M or early-L dwarf at the distance of the primary,
as shown in Figures~\ref{fig:cmd} and \ref{fig:spw2}.
Spectroscopy is needed to confirm that it has a spectral type in this range.

\subsection{WISE J041328.73+821854.7}

\begin{figure}
\epsscale{1}
\plotone{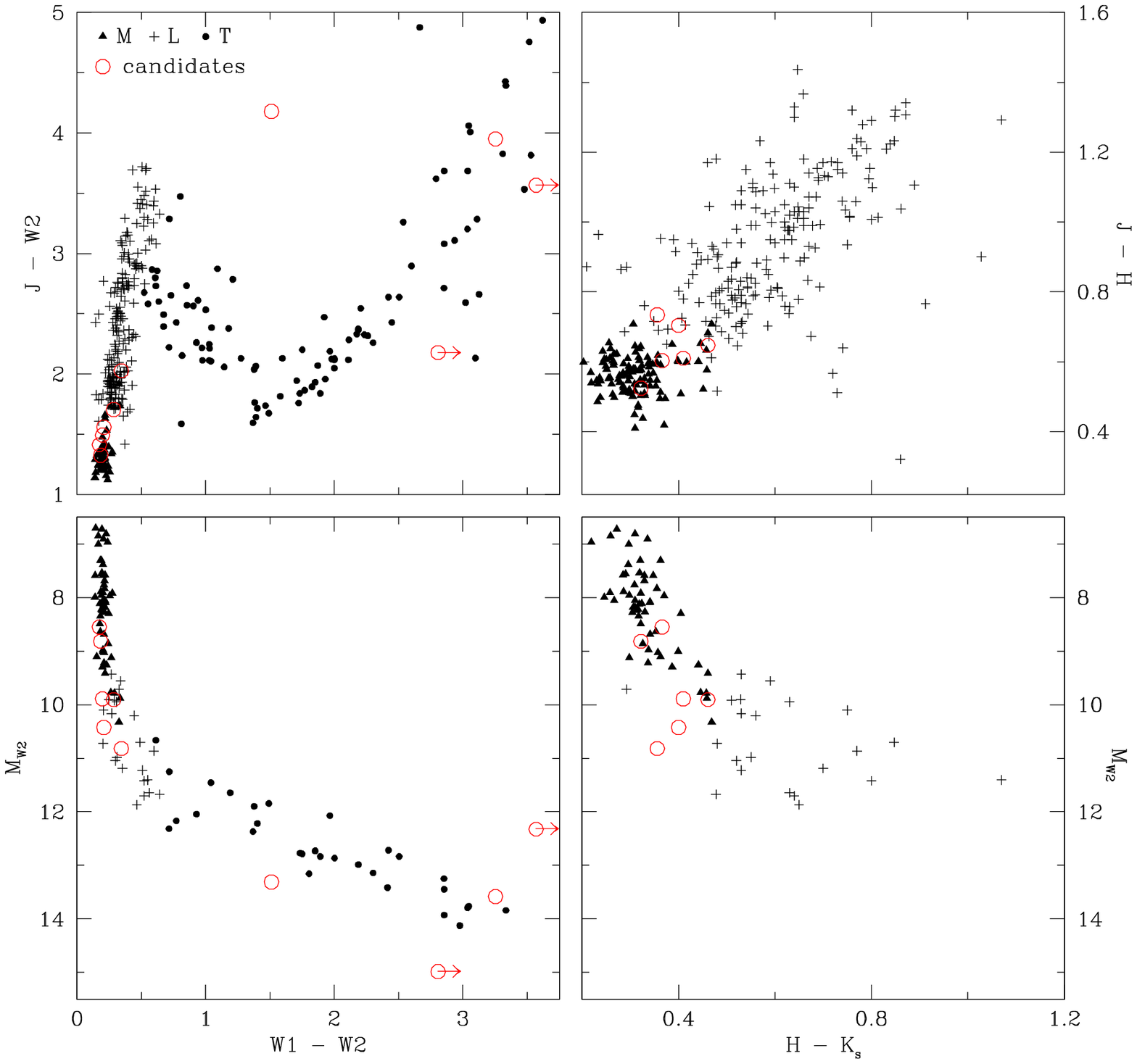}
\caption{
Color-magnitude and color-color diagrams for candidate late-type
companions (open circles) and a sample of M, L, and T dwarfs
\citep[filled triangles, crosses, and
filled circles,][http://DwarfArchives.org]{leg10a,kir11,kir12}.
The T dwarfs are not
shown in the diagrams on the right since our faintest candidates lack
measurements of $K_s$. The candidate companions are assumed to have
the same distances as their primaries for the bottom diagrams.
}
\label{fig:cmd}
\end{figure}

Our examination of {\it WISE} sources lacking 2MASS counterparts uncovered
WISE J041328.73+821854.7 (hereafter WISE 0413+8218) as a candidate companion
to LHS~1643.
\citet{law08} estimated a spectral type of M4.5 for the latter from
a measurement of $V-K$. 
The 2MASS photometry of LHS~1643 would then imply a distance of 22~pc.
If this distance is adopted for WISE~0413+8218, its position in $M_{W2}$
versus $W1-W2$ is consistent with that of a T dwarf, as illustrated
in Figure~\ref{fig:cmd} ($W1-W2=1.5$), and its angular separation of
$215\arcsec$ would correspond to 4700~AU. Because this separation is small
enough that companionship is plausible \citep[$\lesssim$10,000~AU,][]{lep07},
we pursued followup imaging of WISE~0413+8218 to measure its proper motion
and better constrain its spectral type. Using the WIYN High-Resolution
Infrared Camera (WHIRC),
we obtained nine dithered 30~sec exposures of
WISE~0413+8218 through a $J$-band filter on 2011 October 11.
Only a weak detection of WISE~0413+8218 appeared in the reduced mosaic of
these data, but it was sufficient to demonstrate that the candidate
is much redder in $J-W2$ than expected for a T dwarf with its value of $W1-W2$
(Figure~\ref{fig:cmd}). In addition, no motion was detected for WISE~0413+8218 
between the {\it WISE} and WHIRC images, which is marginally inconsistent
with companionship (Table~\ref{tab:comp}). Therefore, we conclude that
WISE~0413+8218 is probably not a companion to LHS~1643.

\subsection{ULAS J095047.28+011734.3}

\begin{figure}
\epsscale{1}
\plotone{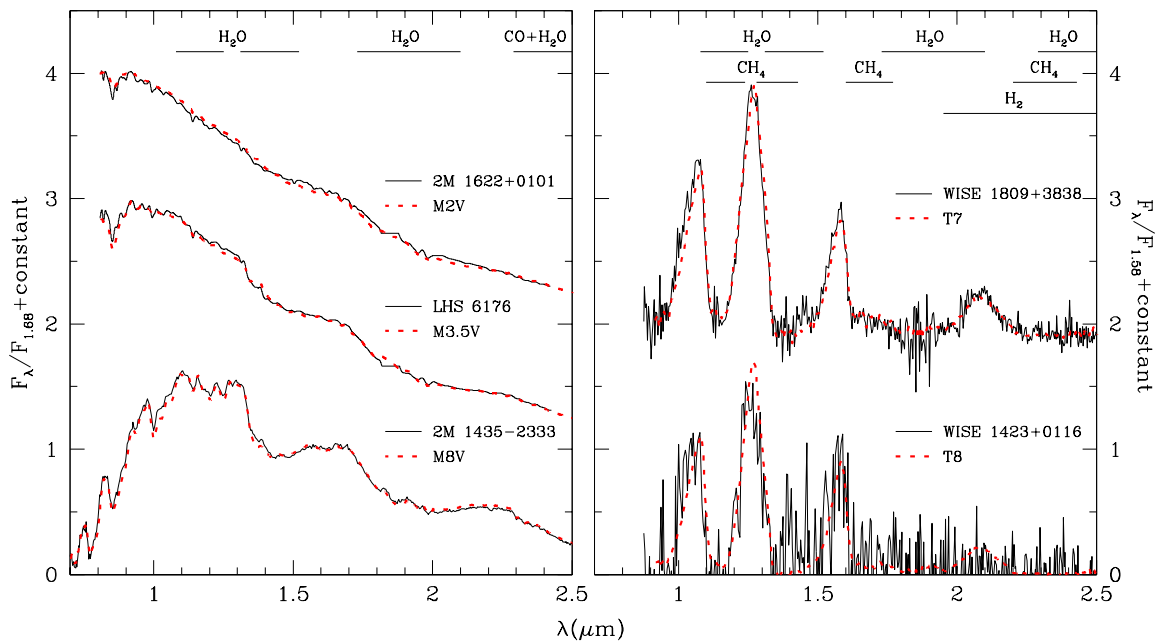}
\caption{
Near-IR spectra of the primary star LHS~6176 and four of the candidate
companions (solid lines) compared to data for dwarf standards 
\citep[dotted lines,][]{cus05,ray09,bur04,bur06}.
The data are displayed at a resolution of $R=150$.
}
\label{fig:spex}
\end{figure}

The primary for our next candidate companion is LHS~6176.
Because this star has not been
previously classified, we performed spectroscopy on it with SpeX \citep{ray03}
at the NASA Infrared Telescope Facility (IRTF). On the night of 2011 April 23,
we observed LHS~6176 in the SXD mode with a $0\farcs8$ slit, which provided a
wavelength coverage of 0.8--2.5~\micron\ and a resolution of $R=750$.
The data were reduced with the Spextool package \citep{cus04} and
corrected for telluric absorption \citep{vac03}. The reduced spectrum
is shown in Figure~\ref{fig:spex}, where it has been smoothed to the resolution
of the SpeX prism data that we have collected for some of our other candidates
($R=150$).  By comparing its spectrum to SpeX data for standard dwarfs
\citep{cus05,ray09}, we measured a spectral type of M3.5V$\pm$1 for LHS~6176.
We estimated a distance of 27~pc based on this classification and
the photometry from 2MASS.
Proper motions have been previously measured for LHS~6176 using DSS images
\citep{bak02,lep05b}. However, it is blended with another star
in the images from the Second Palomar Sky Survey, which may have introduced
errors into those data. Therefore, we have performed a new measurement 
using the first-epoch Palomar Sky Survey and archival {\it Spitzer}
4.5~\micron\ images from 2011 (Table~\ref{tab:comp}).

The candidate companion to LHS~6176 is ULAS~J095047.28+011734.3 (hereafter
ULAS~0950+0117). B. Burningham (in preparation) discovered this object
using UKIDSS data and classified it as a T8 dwarf. We identified it as
a possible companion through two separate surveys. In the first one, we
searched for common proper motion companions using multi-epoch images from
{\it Spitzer} \citep{luh11}. Because ULAS~0950+0117 was observed by
{\it Spitzer} in 2010 (program 60093, S. Leggett), we selected it for
a second epoch of imaging with the satellite in 2011 (program 70021, K. Luhman).
By comparing the astrometry for all objects detected in both epochs,
we found the proper motion of LHS~6176 was similar to that of ULAS~0950+0117.
We later recovered ULAS~0950+0117 as a candidate companion during
our analysis of {\it WISE} sources lacking 2MASS counterparts that are
in close proximity to nearby stars like LHS~6176 (Section~\ref{sec:methods}).

ULAS~0950+0117 and LHS~6176 have an angular separation of $52\arcsec$,
corresponding to 1400~AU at the distance of the latter.
UKIDSS provides the earliest astrometry of ULAS~0950+0117.
It was detected in $H$ and $K$ in 2006 and 2007, but the
signal-to-noise ratios were low. Therefore, we use the $J$-band data
from 2008 February 23 for the first epoch in our measurement of the proper
motion of ULAS~0950+0117.
For the second epoch, we used the most recent publicly available 
images that were taken near the same time of year as the first epoch to
minimize parallactic motion, which consist of $J$-band data acquired
on the night of 2012 March 3 with SOFI
on the European Southern Observatory New Technology Telescope (NTT, program
186.C-0756, R. Smart). The resulting proper motion measurement for
ULAS~0950+0117 is similar to that of LHS~6176 such that their 1~$\sigma$
error bars overlap (Table~\ref{tab:comp}).
The diagram of $M_{W2}$ versus spectral type in Figure~\ref{fig:spw2} shows
that ULAS~0950+0117 is brighter than expected for a T8 dwarf at the
spectrophotometric distance
of LHS~6176, which may indicate that it is an unresolved binary.
ULAS~0950+0117 has been imaged several times over the last few years with
both the NTT and the Canada-France-Hawaii Telescope (CFHT) in an effort
to measure its parallax \citep{mar10,dup12}, which will eventually provide
an additional test of its companionship and overluminous nature.

\subsection{2MASS J10313234$-$5338010}

2MASS~J10313234$-$5338010 (hereafter 2M~1031$-$5338) was found as a possible
companion to HD~91324 (F9V, 21.8~pc) through our 2MASS/{\it WISE} proper
motion survey.
The angular separation of the two stars is $309\arcsec$ (6700~AU).
Our proper motion measurement for 2M~1031$-$5338 from 2MASS and {\it WISE}
is within 1~$\sigma$ of that
of HD~91324 (Table~\ref{tab:comp}). Using Figures~\ref{fig:cmd}
and \ref{fig:spw2}, we find that the photometry of 2M~1031$-$5338 is
indicative of an M5--M6 dwarf near the distance of HD~91324.
As with 2M~0318+0828, spectroscopy is needed to verify this spectral type.

\subsection{WISE J142320.84+011638.0}

We identified WISE~J142320.84+011638.0 (hereafter WISE 1423+0116)
as a candidate companion to LHS~2907 (BD+01$\arcdeg$2920) from among
{\it WISE} sources lacking 2MASS counterparts.
LHS~2907 is a moderately metal-poor solar-type star 
\citep[\lbrack Fe/H\rbrack $\sim-$0.35,][]{val05,cas11}.
The pair has an angular separation of $156\arcsec$, or 2680~AU at the
distance of the primary (17.2~pc).
After checking data archives from various surveys and observatories, we
found detections of WISE~1423+0116 in $Y$- and $J$-band images from UKIDSS
and $J$-band images from WIRCam at the CFHT (program 09BD95, L. Albert).
The values of $W2$, $W1-W2$, $J-W2$ are consistent with
a late T dwarf at the distance of LHS~2907 (Figure~\ref{fig:cmd}).
Using the images from UKIDSS (2008 May 16) and CFHT (2009 December 29),
we measured the proper motion of WISE~1423+0116, which was found to be
similar to that of LHS~2907 (Table~\ref{tab:comp}). 
Because of this promising evidence that WISE~1423+0116 was a T dwarf
companion, we obtained a spectrum of it with SpeX at the IRTF.
These data were collected in the prism mode with a $0\farcs8$ slit, providing
a wavelength coverage of 0.8--2.5~\micron\ and a resolution of $R=150$.
The spectrum exhibits the strong absorption bands that characterize
late T dwarfs, as illustrated in Figure~\ref{fig:spex} where we compare
it data for a standard T8 dwarf \citep[2MASS~J04151954$-$0935066,][]{bur04}.
During their independent discovery of WISE~1423+0116, \citet{pin12}
obtained a spectrum with higher signal-to-noise ratio, producing a more accurate
classification than is possible with our data.
Using their spectral type of T8 and the distance of the primary,
WISE~1423+0116 falls within the sequence of T dwarfs in the diagram
of $M_{W2}$ versus spectral type in Figure~\ref{fig:spw2}, which supports
its companionship.

\subsection{2MASS J14351087$-$2333025}

Our 2MASS/{\it WISE} proper motion search uncovered 2MASS~J14351087$-$2333025
(hereafter 2M~1435$-$2333) as a possible companion to 2MASS J14345819$-$2335572
(hereafter 2M~1434$-$2335).
The latter has a spectral type of M7V and a corresponding
distance of 27~pc. The pair had an angular separation of $247\arcsec$ in
the 2MASS images (6700~AU). The proper motion that we measure for
2M~1435$-$2333 using the first-epoch Palomar Sky Survey, 2MASS, and
{\it WISE} (Table~\ref{tab:comp}) agrees with the value of
($\mu_\alpha$,$\mu_\delta$)=($-306\pm30$,$+51\pm5$)~mas~yr$^{-1}$
for 2M~1434$-$2335 from \citet{fah09}. To further characterize 2M~1435$-$2333,
we obtained a spectrum of it using the prism mode of SpeX on the night
of 2012 May 6. The spectrum matches our SpeX data for the M8V standard
2MASS J03205965+1854233 \citep{kir95}, as shown in Figure~\ref{fig:spex}.
Our classification of M8V$\pm$0.5 is roughly consistent
with companionship given its photometry. However, when we measure the proper
motion of 2M~1434$-$2335 in the same manner as for 2M~1435$-$2333, we arrive
at ($\mu_\alpha$,$\mu_\delta$)=($-$314,$-$19)$\pm$15~mas~yr$^{-1}$, which
differs from the motion from \citet{fah09}. Thus, the motions that we measure
for the components of this pair indicate that they are not companions.

\subsection{2MASS J16220644+0101156}

2MASS J16220644+0101156 (hereafter 2M~1622+0101) is a candidate companion to
HD~147449 (F0V, 27.3~pc) based on our 2MASS/{\it WISE} proper motion analysis.
2M~1622+0101 is relatively close to HD~147449 with an angular separation of
only $43\arcsec$ (1200 AU). The proper motion of 2M~1622+0101
that we measured from 2MASS and {\it WISE} agrees with that of
HD~147449 (Table~\ref{tab:comp}). To measure the spectral
type of 2M~1622+0101, we performed spectroscopy on it with SpeX in its SXD
mode ($R=750$) on the night of 2012 April 24. Through comparison
to SpeX data for standard stars \citep{cus05,ray09}, we classified 2M~1622+0101
as M2V$\pm$1. The spectrum of 2M~1622+0101 is plotted with data for the M2V star
Gl~411 in Figure~\ref{fig:spex}, both of which have been smoothed
to the resolution of the prism spectra ($R=150$). Our classification
is consistent with the expected spectral type based on its
photometry if it has the same distance as HD~147449. Because of its
relatively early spectral type, 2M~1622+0101 does not appear within
the boundaries of Figures~\ref{fig:cmd} and \ref{fig:spw2}.

\subsection{2MASS J17430860+8526594}

Using 2MASS and {\it WISE}, we identified 2MASS J17430860+8526594
(hereafter 2M~1743+8526) as a candidate companion to G259-20.
Neither a spectral type nor a parallax has been measured previously for
G259-20, but its optical and IR photometry are indicative of an early-M dwarf
at 20--30~pc. 
The proper motion of 2M~1743+8526 based on 2MASS and {\it WISE} agrees
with that of G259-20 (Table~\ref{tab:comp}) and the pair has a relatively
small separation ($30\arcsec$, 650~AU). 
In addition, the photometry of 2M~1743+8526 is consistent with that of a
late-type object near the distance of G259-20 (Figure~\ref{fig:cmd}). Because
of the promising nature of this candidate, 
we pursued spectroscopy of it and G259-20 with the Near-Infrared Spectrometer
\citep[NIRSPEC,][]{mcl98,mcl00} on the 10~m telescope at W. M. Keck Observatory 
on the night of 2012 June 7.
The instrument was operated in its low-resolution mode with the N3 filter
and $0\farcs57$ slit, which provides spectra across the $J$ band with a
resolution of $R=1700$. The data were reduced with the REDSPEC package
\citep{mcl03} and the resulting spectra are presented in
Figure~\ref{fig:nirspec}. We measured spectral types of
M2.5V$\pm$1 and L5$\pm$1 for G259-20 and 2M~1743+8526, respectively,
using spectra of standards obtained with SpeX \citep{cus05,ray09} and
NIRSPEC \citep{mcl03}.
Based on the metallicities of the early-M SpeX standards \citep[e.g.,][]{roj12},
we estimate that G259-20 has \lbrack Fe/H\rbrack$\approx-$0.2 to 0,
which is relevant to the interpretation of unusual colors for
2M~1743+8526 (Section~\ref{sec:disc}).
In Figure~\ref{fig:nirspec}, we show the standards that produced
the best matches, consisting of Gl~581 (M2.5V) for the primary and
DENIS-P~J1228.2$-$1547 (L5) for the secondary.
When combined with its 2MASS photometry, a spectral type of M2.5V for G259-20
implies a distance of 22~pc. 2M~1743+8526 is plotted on the diagram of
$M_{W2}$ versus spectral type in Figure~\ref{fig:spw2} assuming this distance,
which places it within the sequence of known L dwarfs, as expected for a
companion.

\subsection{WISE J180901.07+383805.4}

We identified WISE~J180901.07+383805.4 (hereafter WISE~1809+3838) as a
candidate companion to HD~166620 (K2V, 11.0~pc) from among {\it WISE}
sources that do not appear in 2MASS. The angular separation is
$769\arcsec$, or 8460~AU at the distance of HD~166620.
Given its {\it WISE} photometry ($W1-W2>2.8$), WISE~1809+3838 could
be a companion to HD~166620 with a spectral type of $>$T9, as
shown in Figure~\ref{fig:cmd}.
We found a detection of WISE~1809+3838 in archival $J$-band images obtained
by WIRCam at the CFHT (program 08BD95, L. Albert) on the night of 2008
October 12. 
The photometry that we measured from these data indicates
a color of $J-W2=2.2$, which is bluer than expected for a $>$T9 companion.
Nevertheless, to investigate the nature of WISE~1809+3838 further,
we observed it with SpeX in the prism mode on the night of 2011 June 28.
The resulting spectrum is presented in Figure~\ref{fig:spex}, where we
find a good match to the T7 standard 2MASS~J07271824+1710012 \citep{bur06}.
The uncertainty in this classification is $\pm1$~subclass.
If WISE~1809+3838 has an absolute magnitude in W2 that is similar to that
of typical T7 dwarfs, than it should have a distance of roughly 28~pc,
indicating that it is probably not a companion to HD~166620.
In addition to the spectroscopic data, we obtained $J$-band images of
WISE~1809+3838 with the slit-viewing camera on SpeX.
The proper motion of WISE~1809+3838 measured from the SpeX and WIRCam
data is inconsistent with that of HD~166620 (Table~\ref{tab:comp}),
confirming that the pair is not a binary system.

\subsection{2MASS J18525777$-$5708141}

2MASS~J18525777$-$5708141 (hereafter 2M~1852$-$5708) was found as
a candidate companion to LHS~3421 (M2.5V, 25.9~pc) through
our 2MASS/{\it WISE} proper motion search. 
The pair has an angular separation of $58\arcsec$ (1500~AU).
The proper motion of 2M~1031$-$5338 based on 2MASS and {\it WISE}
is within 1~$\sigma$ of that of LHS~3421 (Table~\ref{tab:comp}).
The colors and magnitudes of 2M~1852$-$5708 are consistent
with those of a dwarf between M5--M6 at the distance of the primary
(Figs.~\ref{fig:cmd} and \ref{fig:spw2}).

\subsection{2MASS J20103539+0634367}

\begin{figure}
\epsscale{1}
\plotone{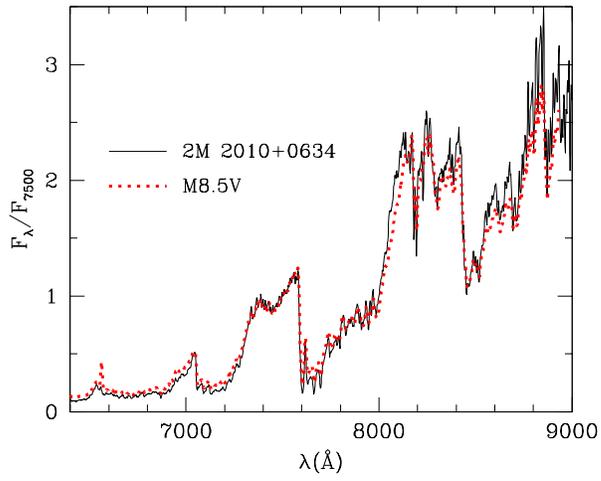}
\caption{
Optical spectrum of the candidate companion 2M~2010+0634 (solid line)
compared to the average of data for M8V and M9V standards (dotted line).
}
\label{fig:het}
\end{figure}

We uncovered 2MASS~J20103539+0634367 (hereafter 2M~2010+0634) as
a candidate companion to the M-type spectroscopic binary LSPM~J2010+0632
with data from 2MASS and {\it WISE}.
The latter has a distance of 15.3~pc based on its spectral type and
photometry \citep{shk10}.
The angular separation of the pair is $143\arcsec$ (2100~AU).
The proper motion of 2M~2010+0634 derived from 2MASS and {\it WISE}
agrees with that of LSPM~J2010+0632 (Table~\ref{tab:comp}).
The photometry of 2M~2010+0634 also is consistent with a spectral type of
late M or early L at the distance of the primary (Figure~\ref{fig:cmd}).
To verify its spectral type, we obtained a spectrum of 2M~2010+0634 with the
Marcario Low-Resolution Spectrograph (LRS) on the Hobby-Eberly Telescope (HET)
on the night of 2011 August 26. The instrument was operated with the G3 grism
and the $2\arcsec$ slit, which provided a wavelength coverage of
6200--9100~\AA\ and a resolution of $R=1100$.
As illustrated in Figure~\ref{fig:het},
the spectrum agrees well with a combined spectrum of
VB~10 (M8V), LHS~2243 (M8V), and LHS~2243 (M9V) in which the latter
was given twice the weighting of each of the other two stars.
Therefore, we classified 2M~2010+0634 as M8.5V$\pm$0.5.
Using this spectral type and the distance of the primary,
2M~2010+0634 appears within the sequence of M dwarfs in
Figure~\ref{fig:spw2}, which supports its companionship.

\section{Discussion}
\label{sec:disc}

We have presented 11 candidate late-type companions to nearby stars
that were found with two methods involving data from {\it WISE} and 2MASS.
In general, the probability of companionship for a candidate depends
on the size of its proper motion ($\mu$), the differences between the proper
motions and distances of the candidate and the primary ($\Delta\mu$,
$\Delta d$), the uncertainties in those motions and distances ($\sigma_\mu$,
$\sigma_d$), and the angular separation of the pair ($\Delta\theta$).
For eight of our candidates, the values of these parameters are comparable
to those of cool dwarfs that have been previously accepted as companions. 
For instance, these objects satisfy by wide margins the criteria for
companionship proposed
by \citet[][$\Delta\theta\Delta\mu<(\mu/0.15)^{3.8}$]{lep07}
and \citet[][$\Delta\mu/\mu<0.2$]{dup12}.
These likely companions, which are indicated in Table~\ref{tab:comp},
include two M dwarfs, one L dwarf, and two T dwarfs.
The three remaining companions lack spectroscopic classifications, but they
probably have M/L types based on their photometry.

Cooler companions have greater value for testing model atmospheres and
interiors. 
The T dwarf companion WISE~1423+0116 was independently discovered and previously
reported by \citet{pin12}, who thoroughly examined its physical properties.
Our remaining L/T companions are ULAS~0950+0117 and 2M~1743+8526.
Because their primaries are not well-studied (e.g., lack parallax data),
it is not currently possible to accurately estimate their
properties and fully exploit them for calibrating models.
However, both companions exhibit distinctive characteristics that warrant
discussion. If it is a companion, ULAS~0950+0117 is brighter than
typical T dwarfs at its spectral type according to the primary's
spectrophotometric distance, indicating that it may be an unresolved
binary. Thus, estimating the properties of this object for comparison to
models will require better constraints on its binarity via high-resolution
imaging. If confirmed as a binary, ULAS~0950+0117 would join a small group of
known L and T dwarf binaries in orbit around stars
\citep{wil01,pot02,bou03,mcc04,bur05}. Such pairs are especially useful for
testing theoretical models since their ages and metallicities can be estimated
via the stellar primaries while their masses can be measured dynamically
\citep{dup09,kon10}.

\begin{figure}
\epsscale{1}
\plotone{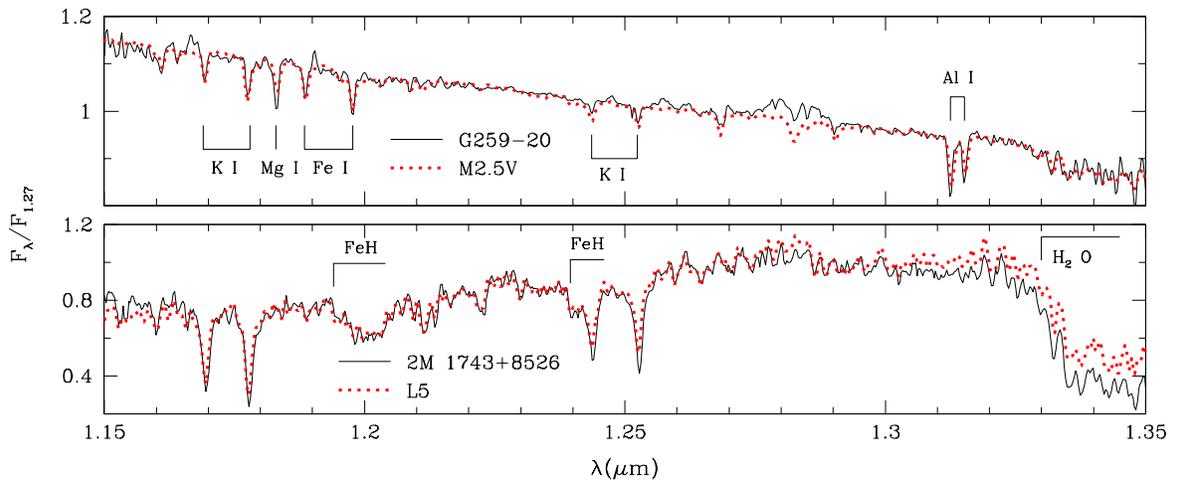}
\caption{
Near-IR spectra of the primary star G259-20 and its candidate companion
2M~1743+8526 (solid lines) compared to data for dwarf standards
(dotted lines).
}
\label{fig:nirspec}
\end{figure}

2M~1743+8526 is also notable in that it has unusually blue near-IR colors for
its spectral type ($J-K_s=1.09\pm0.06$). A number of blue L dwarfs have been
previously identified \citep[][references therein]{cru03,kna04,kir10}. 
Two of the most prominent examples, 2MASS~J11263991$-$5003550
\citep{bur08} and SDSS~J141624.08+134826.7 \citep[hereafter
SDSS~1416+1348,][]{bow10,schm10},
resemble 2M~1743+8526 in terms of colors and spectral type.
The 1.15--1.3~\micron\ spectral features of 2M~1743+8526 match those
of normal L5 dwarfs, but it has stronger H$_2$O absorption at
1.33--1.35~\micron\ (Figure~\ref{fig:nirspec}), which is similar to the
behavior of SDSS~1416+1348 \citep{schm10}.
A comparison between 2M~1743+8526 and these two L dwarfs at other wavelengths
would be useful for determining whether they share other spectral
characteristics as well, although observations of the former are
hampered by its declination, which places it at high airmass and
above the declination limits of some telescopes.

Possible explanations for the peculiar colors of blue L dwarfs
include low metallicity, unresolved companions, and condensate clouds that are
thin or have large grains.
Most studies have favored the latter \citep{bur08,bow10,cus10}, but
the properties of the T dwarf companion to SDSS~1416+1348
suggest that low metallicity plays a role as well \citep{dup12}.
Because some of its properties can be estimated through its primary star,
2M~1743+8526 offers a new opportunity for constraining these proposed
mechanisms.
Adopting the spectrophotometric distance of its primary, the absolute
magnitudes of 2M~1743+8526 are brighter than the average values for L5
dwarfs \citep{dup12} by 0.72 ($J$), 0.38 ($H$), 0.17 ($K_s$), 0.01 (W1),
and 0.03~mag (W2). The enhanced $J$-band flux is consistent with
the third scenario described above \citep{bur08b}.
Meanwhile, our metallicity estimate for the primary
(\lbrack Fe/H\rbrack$\approx-$0.2 to 0) indicates that it is not an outlier
in the metallicity distribution in the solar neighborhood \citep{nor04},
and hence low metallicity
is probably not the cause of 2M~1743+8526's unusual colors. \citet{rad08}
arrived at a similar result through analysis of the primary for the blue L5 
companion 2MASS J17114559+4028578 ($J-K_s=1.21\pm0.08$). Thus, the data for
2M~1743+8526 support the previous suggestion that thin condensate clouds
can account for some blue L dwarfs.

\acknowledgements
K. L. acknowledges support from grant AST-0544588
from the National Science Foundation (NSF) and grant NNX12AI47G 
from the NASA Astrophysics Data Analysis Program.
\wise is a joint project of the University of California, Los Angeles,
and the Jet Propulsion Laboratory (JPL)/California Institute of
Technology (Caltech), funded by NASA. 
2MASS is a joint project of the University of
Massachusetts and the Infrared Processing and Analysis Center (IPAC) at
Caltech, funded by NASA and the NSF.
The W.M. Keck Observatory is operated as a scientific partnership among
Caltech, the University of California, and NASA and was made possible by
the generous financial support of the W.M. Keck Foundation. 
The IRTF is operated by the University of Hawaii under cooperative agreement
NNX-08AE38A with NASA.
The HET is a joint project of the University of Texas at Austin, the
Pennsylvania State University, Stanford University,
Ludwig-Maximillians-Universit\"at M\"unchen, and Georg-August-Universit\"at
G\"ottingen and is named in honor of its principal benefactors, William
P. Hobby and Robert E. Eberly. The {\it Spitzer Space Telescope}
is operated by JPL and Caltech under contract with NASA.
Kitt Peak National Observatory is operated by the Association of Universities
for Research in Astronomy, Inc. under cooperative agreement with the NSF.
WIYN Observatory is a joint facility of the University of Wisconsin-Madison,
Indiana University, Yale University, and the National Optical Astronomy
Observatory.
WIRCam is a joint project of CFHT, Taiwan, Korea, Canada, and France.
CFHT is operated by the National Research Council of Canada, the Institute
National des Sciences de l'Univers of the Centre National de la Recherche
Scientifique of France, and the University of Hawaii.
This work uses data from the SpeX Prism Spectral Libraries 
(maintained by Adam Burgasser at http://www.browndwarfs.org/spexprism),
the M, L, and T dwarf compendium at http://DwarfArchives.org
(maintained by Chris Gelino, Davy Kirkpatrick, and Adam Burgasser),
the Brown Dwarf Spectroscopic Survey Archive
(http://www.astro.ucla.edu/$\sim$mclean/BDSSarchive),
the NASA/IPAC Infrared Science Archive (operated by JPL
under contract with NASA), and the ESO/ST-ECF Science Archive Facility.
The Center for Exoplanets and Habitable Worlds is supported by the
Pennsylvania State University, the Eberly College of Science, and the
Pennsylvania Space Grant Consortium.
The authors wish to recognize and acknowledge the very significant cultural
role and reverence that the summit of Mauna Kea has always had within the
indigenous Hawaiian community. We are most fortunate to have the opportunity
to conduct observations from this mountain.

\clearpage

\begin{deluxetable}{llllllllll}
\rotate
\tabletypesize{\scriptsize}
\tablewidth{0pt}
\tablecaption{Spectral Types, Distances, and Astrometry for Candidate Binary Systems\label{tab:comp}}
\tablehead{
\colhead{Primary} & \colhead{Spectral} & \colhead{Distance} & \colhead{$\mu_{\alpha}$,$\mu_{\delta}$\tablenotemark{a}} & \colhead{Ref} &
\colhead{Candidate} & \colhead{Separation} & \colhead{Spectral} 
& \colhead{$\mu_{\alpha}$,$\mu_{\delta}$\tablenotemark{a,b}} & \colhead{Likely} \\
\colhead{} & \colhead{Type} & \colhead{(pc)} & \colhead{(mas~yr$^{-1}$)} &
\colhead{} & \colhead{Companion} & \colhead{(arcsec/AU)} & \colhead{Type} & 
\colhead{(mas~yr$^{-1}$)} & \colhead{Binary?}}
\startdata
LSPM J0318+0827S & M3V & 23  & ($-$234,$-$167)$\pm$8 & 1,2,3 & 2MASS J03184214+0828002 & 75/1700 & \nodata  & ($-$252,$-$174)$\pm$23 & yes \\
LHS 1643 & mid-M? & 22  & (+260,$-$573)$\pm$8 & 4,2,3 & WISE J041328.73+821854.7 & 215/4700 & \nodata  & ($<$300,$<$300) & no \\
LHS 6176 & M3.5V & 27  & (+241,$-$359)$\pm$13 & 2 & ULAS J095047.28+011734.3 & 52/1400 & T8\tablenotemark{c}  & (+245,$-$380)$\pm$13 & yes \\
HD 91324 & F9V & 21.8  & ($-$419.3,+209.2)$\pm$0.2 & 5,6 & 2MASS J10313234$-$5338010 & 309/6700 & \nodata & ($-$400,+211)$\pm$23 & yes \\
LHS 2907 & G1V & 17.2 & (+223.8,$-$477.4)$\pm$0.4 & 7,6 & WISE J142320.84+011638.0 & 156/2680\tablenotemark{d} & T8p\tablenotemark{e} & (+280,$-$420)$\pm$60 & yes \\
2MASS J14345819$-$2335572 & M7V & 27 & ($-$314,$-$19)$\pm$15 & 8,2 & 2MASS J14351087$-$2333025 & 247/6700 & M8V & ($-$310,+54)$\pm$15 & no \\ 
HD 147449 & F0V & 27.3  & ($-$158.4,+49.6)$\pm$0.4 & 9,6 & 2MASS J16220644+0101156 & 43/1200 & M2V & ($-$155,+55)$\pm$35 & yes \\
G259-20 & M2.5V & 22 & ($-$104.1,$-$269.5)$\pm$2.7 & 2,10 & 2MASS J17430860+8526594 & 30/650 & L5 & ($-$94,$-$271)$\pm$25 & yes \\
HD 166620 & K2V & 11.0  & ($-$316.4,$-$468.5)$\pm$0.3 & 7,6 & WISE J180901.07+383805.4 & 769/8460 & T7 & ($-$555,$-$48)$\pm$60 & no \\
LHS 3421 & M2.5V & 25.9 & ($-$246,$-$769)$\pm$4 & 11,6 & 2MASS J18525777$-$5708141 & 58/1500 & \nodata & ($-$242,$-$743)$\pm$35 & yes \\
LSPM J2010+0632 & M3.5V & 15 & (+43,$-$195)$\pm$8 & 12,3 & 2MASS J20103539+0634367 & 143/2100 & M8.5V & (+51,$-$204)$\pm$20 & yes \\
\enddata
\tablenotetext{a}{Absolute proper motions are listed for HD~91324, LHS~2907,
HD~147449, G259-20, HD~166620, and LHS~3421. All other proper motions are
relative to surrounding background stars.}
\tablenotetext{b}{This work.}
\tablenotetext{c}{B. Burningham, in preparation.}
\tablenotetext{d}{\citet{pin12} reported a separation of 
$153\arcsec$, which appears to be based on coordinates for the primary 
and secondary that are from different epochs.}.
\tablenotetext{e}{\citet{pin12}.}
\tablerefs{
(1) \citet{rob84};
(2) this work;
(3) \citet{lep05b};
(4) \citet{law08};
(5) \citet{gra06};
(6) \citet{van07};
(7) \citet{cow67};
(8) \citet{cru07};
(9) \citet{sle55};
(10) \citet{hog00};
(11) \citet{haw96};
(12) \citet{shk10}.
}
\end{deluxetable}

\clearpage

\begin{deluxetable}{lllllllll}
\tabletypesize{\scriptsize}
\tablewidth{0pt}
\tablecaption{Photometry for Candidate Companions\label{tab:phot}}
\tablehead{
\colhead{Name} & \colhead{$J$} & \colhead{$H$} & \colhead{$K_s$} & \colhead{Ref} & \colhead{W1} & \colhead{W2} & \colhead{W3} & \colhead{W4}}
\startdata
2MASS 03184214+0828002 & 13.79$\pm$0.03 & 13.09$\pm$0.03 & 12.69$\pm$0.02 & 1 & 12.44$\pm$0.02 & 12.23$\pm$0.03 & 11.42$\pm$0.19 & \nodata \\
WISE J041328.73+821854.7 & $\sim$19.2 & \nodata & \nodata & 2 & 16.53$\pm$0.10 & 15.02$\pm$0.09 & 12.43$\pm$0.32 & \nodata \\
ULAS J095047.28+011734.3 & 18.05$\pm$0.04 & 18.24$\pm$0.15 & \nodata & 3 & $<$18\tablenotemark{a} & 14.48$\pm$0.06 & \nodata & \nodata \\
2MASS J10313234$-$5338010 & 11.84$\pm$0.02 & 11.31$\pm$0.02 & 10.99$\pm$0.02 & 1 & 10.70$\pm$0.02 & 10.51$\pm$0.02 & 10.88$\pm$0.19 & \nodata \\
WISE J142320.84+011638.0 & 18.71$\pm$0.05 & 19.14$\pm$0.20 & \nodata & 4 & 18.01$\pm$0.30 & 14.85$\pm$0.07 & \nodata  & \nodata \\
2MASS J14351087$-$2333025 & 13.54$\pm$0.02 & 12.93$\pm$0.03 & 12.52$\pm$0.03 & 1 & 12.25$\pm$0.02 & 12.05$\pm$02 & \nodata & \nodata \\
2MASS J16220644+0101156 & 8.68$\pm$0.03 & 8.00$\pm$0.02 & 7.82$\pm$0.02 & 1 & 7.69$\pm$0.02 & 7.62$\pm$0.02 & 7.54$\pm$0.02 & 7.39$\pm$0.11 \\
2MASS J17430860+8526594 & 14.56$\pm$0.04 & 13.82$\pm$0.04 & 13.47$\pm$0.05 & 1 & 12.88$\pm$0.03 & 12.53$\pm$0.03 & $<$12.4\tablenotemark{a} & \nodata \\
WISE J180901.07+383805.4 & 17.37$\pm$0.05 & \nodata & \nodata & 2 & $<$18 & 15.19$\pm$0.09 & \nodata & \nodata \\
2MASS J18525777$-$5708141 & 12.03$\pm$0.02 & 11.43$\pm$0.02 & 11.06$\pm$0.02 & 1 & 10.79$\pm$0.02 & 10.62$\pm$0.02 & 10.40$\pm$0.06 & \nodata \\
2MASS J20103539+0634367 & 12.53$\pm$0.02 & 11.89$\pm$0.02 & 11.43$\pm$0.02 & 1 & 11.11$\pm$0.03 & 10.83$\pm$0.02 & 10.40$\pm$0.06 & \nodata \\
\enddata
\tablecomments{W1--W4 are from the {\it WISE} All-Sky Source Catalog.}
\tablenotetext{a}{A detection is present in the {\it WISE} All-Sky Source
Catalog, but none is apparent from visual inspection of the image.}
\tablerefs{
(1) 2MASS Point Source Catalog;
(2) this work;
(3) UKIDSS Data Release 8;
(4) \citet{pin12}.
}
\end{deluxetable}

\end{document}